\documentclass[preprint,amsmath,amssymb,11pt,english,aps,showpacs,showkeys]{revtex4}
\usepackage{graphicx}
\usepackage{graphics}
\usepackage{amsmath}
\usepackage{dcolumn}
\usepackage{amssymb}
\usepackage{bm}
\usepackage[latin1]{inputenc}

\begin{document}
\title{Noninertial effects on the Dirac oscillator in a topological defect spacetime}
\author{K. Bakke}
\email{kbakke@fisica.ufpb.br}
\affiliation{Departamento de F\'isica, Universidade Federal da Para\'iba, Caixa Postal 5008, 58051-970, Jo\~ao Pessoa, PB, Brazil.}

\begin{abstract}
In this paper, we study the influence of noninertial effects on the Dirac oscillator in the cosmic string spacetime background. We discuss the behaviour of the oscillator frequency in a noninertial system that allows us to obtain relativistic bound state solutions. We also discuss the influence of the topology of the cosmic string spacetime on the relativistic energy levels, and obtain the Dirac spinors for positive-energy solutions. Furthermore, by taking the nonrelativistic limit of the energy levels, we compare the nonrelativistic energy levels to the confinement of a spin-half particle to quantum dot described by the Tan-Inkson model for a quantum dot [W.-C. Tan and J. C. Inkson, Semicond. Sci. Technol. {\bf11}, 1635 (1996)], and a hard-wall confining potential [E. Tsitsishvili {\it et al.}, Phys. Rev. B {\bf70}, 115316 (2004)].
\end{abstract}


\maketitle

\section{Introduction}

In recent years, the Dirac oscillator \cite{osc1} has attracted a great deal of attention in studies of the Ramsey-interferometry effect \cite{osc6}, the quantum Hall effect \cite{osc3}, and hidden supersymmetry \cite{moreno,benitez,quesne2}. The term ``Dirac oscillator" was denominated by Moshinsky and Szczepaniak \cite{osc1} in the study of a relativistic harmonic oscillator based on the introduction of a coupling in such a way that the Dirac equation remains linear in both spatial coordinates and momenta. Moreover, the introduction of this coupling recovers the Schr\"odinger equation for a harmonic oscillator in the nonrelativistic limit of the Dirac equation having a strong spin-orbit coupling. Hence, the Dirac oscillator is given by
\begin{eqnarray}
\vec{p}\rightarrow\vec{p}-im\omega_{0}\rho\,\hat{\beta}\,\hat{\rho},
\label{1}
\end{eqnarray}
where $m$ is the mass of the Dirac neutral particle, $\omega_{0}$ is the oscillator frequency, $\hat{\beta}$ is one of the standard Dirac matrices, and $\hat{\rho}$ is a unit vector on the radial direction. It is worth mentioning that the interaction between a spin-$1/2$ particle and this linear coupling was first investigated by It\^o {\it et al.} \cite{osc13}. Recently, the Dirac oscillator has been investigated in $\left(2+1\right)$ dimensions \cite{osc2,osc8}, in the presence of an external magnetic field \cite{osc9}, in the point of view of the Lie algebra \cite{quesnemo}, by using the shape-invariant method \cite{quesne}, conformal invariance properties \cite{romerom}, in the presence of the Aharonov-Bohm quantum flux \cite{ferk}, and in a system of a charged particle interacting with a topological defect \cite{josevi}.

The aim of this work is to study the influence of noninertial effects on the Dirac oscillator in the background of the cosmic string spacetime. Well-known quantum effects related to noninertial effects are the Sagnac effect \cite{sag,sag5}, the Mashhoon effect \cite{r3}, and the Page-Werner {\it et al.} coupling \cite{r1,r2,r4}. In recent decades, studies of the influence of noninertial effects on quantum systems have been extended to Berry's phase \cite{r7}, scalar fields \cite{r8}, Dirac fields \cite{r10}, quantum interferometry under the influence of gravitational effects \cite{r9}, Lorentz transformations \cite{r5}, the weak field approximation \cite{r6}, the analogue effect of the Aharonov-Casher effect \cite{bf14}, and the Landau-He-McKellar-Wilkens quantization \cite{b}. Recently, the confinement of a neutral particle interacting with external fields to a two-dimensional quantum dot has been achieved via noninertial effects \cite{b2,b4}. Therefore, by choosing a noninertial reference frame, we add a new discussion about the behaviour of the oscillator frequency which allows us to obtain relativistic bound state solutions in the cosmic string spacetime that fills a gap in the studies the Dirac oscillator.

This paper is organized as follows: in section II, we introduce the background of the cosmic string spacetime, and discuss the influence of noninertial effects on the Dirac oscillator by using the spinor theory in curved spacetime. In the following, we discuss the behaviour of the oscillator frequency based on the influence of noninertial effects and the topology of the cosmic string spacetime in order to obtain the relativistic energy levels. We also compare the nonrelativistic limit of the energy levels to the confinement of a spin-half particle to quantum dot described by the Tan-Inkson model for a quantum dot \cite{tan}, and by a hard-wall confining potential \cite{dot,dot2,dot3,bf20}; in section III, we present our conclusions.

\section{Dirac oscillator in the Fermi-Walker reference frame}

In this section, we discuss the influence of the noninertial effects of the Fermi-Walker reference frame on the Dirac oscillator in the cosmic string spacetime. We begin this section by writing the line element of the cosmic string spacetime, and in the following, we build the Fermi-Walker reference frame. The line element of the cosmic string spacetime is characterized by the presence of a parameter related to the deficit of angle which is defined as $\eta=1-4\varpi$, with $\varpi$ being the linear mass density of the cosmic string, and the azimuthal angle being defined in the range: $0\leq\varphi<2\pi$. Working with the units $\hbar=c=1$, the line element of the cosmic string spacetime is given by \cite{vil,kibble}
\begin{eqnarray}
ds^{2}=-dT^{2}+dR^{2}+\eta^{2}R^{2}d\Phi^{2}+dZ^{2}.
\label{2.1}
\end{eqnarray}
Further, the geometry described by the line element (\ref{2.1}) possesses a conical singularity represented by the following curvature tensor $R_{\rho,\varphi}^{\rho,\varphi}=\frac{1-\eta}{4\eta}\,\delta_{2}(\vec{r})$, where $\delta_{2}(\vec{r})$ is the two-dimensional delta function. This behavior of the curvature tensor is denominated as a conical singularity \cite{staro} which gives rise to the curvature concentrated on the cosmic string axis, in all other points the curvature is zero. Moreover, values of the parameter $\eta>1$ correspond to a spacetime with negative curvature which does not make sense in the general relativity context \cite{kleinert,kat,furt}. Hence, the parameter $\eta$ given in the line element (\ref{2.1}) can assume only values for which $\eta<1$. Now, let us make the following coordinate transformation: $T=t$, $R=\rho$, $\Phi=\varphi+\omega\,t$ and $Z=z$,  
where $\omega$ is the constant angular velocity of the rotating frame. Thus, the line element (\ref{2.1}) becomes
\begin{eqnarray}
ds^{2}=-\left(1-\omega^{2}\eta^{2}\rho^{2}\right)\,dt^{2}+2\omega\eta^{2}\rho^{2}d\varphi\,dt+d\rho^{2}+\eta^{2}\rho^{2}d\varphi^{2}+dz^{2}.
\label{2.3}
\end{eqnarray}

We should note that the line element (\ref{2.3}) is defined in the range $0\,<\rho\,<\,1/\omega\eta$. For values where $\rho\,>\,1/\omega\eta$, we can see that the line element (\ref{2.3}) is not defined anymore \cite{landau3}. For $\rho\,>\,1/\omega\eta$, it means that the velocity of the particle is greater than the velocity of the light, thus, the particle would be placed outside of the light-cone. The interest in this restriction on the radial coordinate imposed by noninertial effects is that it gives rise to a hard-wall confining potential \cite{dot}, that is, it imposes that the wave function of the Dirac particle must vanish at $\rho\rightarrow1/\omega\eta$.

Since this topological defect spacetime has a non-null curvature concentrated on the symmetry axis of the cosmic string, we work out the Dirac spinor by using the formulation of spinors in curved spacetime \cite{weinberg}. In a curved spacetime background, spinors are defined in the local reference frame for the observers \cite{weinberg}, where each spinor transform according infinitesimal Lorentz transformations, that is, $\psi'\left(x\right)=D\left(\Lambda\left(x\right)\right)\,\psi\left(x\right)$, where $D\left(\Lambda\left(x\right)\right)$ corresponds to the spinor representation of the infinitesimal Lorentz group and $\Lambda\left(x\right)$ corresponds to the local Lorentz transformations \cite{weinberg}. The local reference frame of the observers can be built through a non-coordinate basis $\hat{\theta}^{a}=e^{a}_{\,\,\,\mu}\left(x\right)\,dx^{\mu}$, whose components $e^{a}_{\,\,\,\mu}\left(x\right)$ are called \textit{tetrads} and satisfy the relation $g_{\mu\nu}\left(x\right)=e^{a}_{\,\,\,\mu}\left(x\right)\,e^{b}_{\,\,\,\nu}\left(x\right)\,\eta_{ab}$ \cite{misner,weinberg,naka}, 
where $\eta_{ab}=\mathrm{diag}(- + + +)$ is the Minkowski tensor. The tetrads have an inverse, $dx^{\mu}=e^{\mu}_{\,\,\,a}\left(x\right)\,\hat{\theta}^{a}$, where we have the relations $e^{a}_{\,\,\,\mu}\left(x\right)\,e^{\mu}_{\,\,\,b}\left(x\right)=\delta^{a}_{\,\,\,b}$ and $e^{\mu}_{\,\,\,a}\left(x\right)\,e^{a}_{\,\,\,\nu}\left(x\right)=\delta^{\mu}_{\,\,\,\nu}$ being satisfied.

We want to build a nonrotating frame called Fermi-Walker reference frame \cite{misner} in order to observe noninertial effects due to the action of external forces without any effects from arbitrary rotations of the local spatial axis of the reference frame of the observers. A Fermi-Walker reference frame \cite{misner} can be built with the components of the non-coordinate basis given in the rest frame of the observers at each instant, that is, $\hat{\theta}^{0}=e^{0}_{\,\,\,t}\left(x\right)\,dt$, and where the spatial components of the non-coordinate basis $\hat{\theta}^{i}$, $i=1,2,3$, do not rotate. Hence, the corresponding local reference frame can be written in the form:
\begin{eqnarray}
\hat{\theta}^{0}=dt;\,\,\,\hat{\theta}^{1}=d\rho;\,\,\,\hat{\theta}^{2}=\eta\omega\rho\,dt+\eta\rho\,d\varphi;\,\,\,\hat{\theta}^{3}=dz.
\label{2.6}
\end{eqnarray}

In order to write the Dirac equation in this curved spacetime background, we need to taking into account that the partial derivative becomes the covariant derivative, where the covariant derivative is given by $\partial_{\mu}\rightarrow\nabla_{\mu}=\partial_{\mu}+\Gamma_{\mu}\left(x\right)$, with $\Gamma_{\mu}\left(x\right)=\frac{i}{4}\,\omega_{\mu ab}\left(x\right)\,\Sigma^{ab}$ being the spinorial connection \cite{bd,naka}, and $\Sigma^{ab}=\frac{i}{2}\left[\gamma^{a},\gamma^{b}\right]$. The indices $(a,b,c=0,1,2,3)$ indicate the local reference frame. The $\gamma^{a}$ matrices are defined in the local reference frame and correspond to the Dirac matrices in the Minkowski spacetime \cite{bd,bjd}:
\begin{eqnarray}
\gamma^{0}=\hat{\beta}=\left(
\begin{array}{cc}
1 & 0 \\
0 & -1 \\
\end{array}\right);\,\,\,\,\,\,
\gamma^{i}=\hat{\beta}\,\hat{\alpha}^{i}=\left(
\begin{array}{cc}
 0 & \sigma^{i} \\
-\sigma^{i} & 0 \\
\end{array}\right);\,\,\,\,\,\,\Sigma^{i}=\left(
\begin{array}{cc}
\sigma^{i} & 0 \\
0 & \sigma^{i} \\	
\end{array}\right),
\label{2.7}
\end{eqnarray}
with $\gamma^{a}\gamma^{b}+\gamma^{b}\gamma^{a}=-2\eta^{ab}$, $\vec{\Sigma}$ being the spin vector, and $\sigma^{i}$ the Pauli matrices. By including the minimal coupling (\ref{1}) that describes the Dirac oscillator, thus, the covariant derivative of a spinor becomes $i\,\nabla_{\mu}\rightarrow i\,\partial_{\mu}+i\,\Gamma_{\mu}\left(x\right)+im\omega_{0}\rho\,\gamma^{0}\,\delta^{\rho}_{\mu}$, 
and the Dirac equation becomes
\begin{eqnarray}
m\psi=i\gamma^{\mu}\,\partial_{\mu}\psi+i\gamma^{\mu}\,\Gamma_{\mu}\left(x\right)\psi+i\gamma^{\mu}\,m\omega_{0}\rho\,\gamma^{0}\,\delta^{\rho}_{\mu}\,\psi.
\label{2.9}
\end{eqnarray}

By using the Fermi-Walker reference frame (\ref{2.6}), we can solve the Maurer-Cartan structure equations in the absence of torsion, that is, $d\hat{\theta}^{a}+\omega^{a}_{\,\,\,b}\wedge\hat{\theta}^{b}=0$, where $\omega^{a}_{\,\,\,b}=\omega_{\mu\,\,\,\,b}^{\,\,\,a}\left(x\right)\,dx^{\mu}$ is the connection 1-form, and obtain the following non-null components of the connection 1-form: $\omega_{\varphi\,\,\,2}^{\,\,\,1}\left(x\right)=-\omega_{\varphi\,\,\,1}^{\,\,\,2}\left(x\right)=-\eta$ and $\omega_{t\,\,\,2}^{\,\,\,1}\left(x\right)=-\omega_{t\,\,\,1}^{\,\,\,2}\left(x\right)=-\omega\eta$.
With these non-null components of the connection 1-form $\omega_{\mu\,\,\,\,b}^{\,\,\,a}\left(x\right)$, we can calculate the components of the spinorial connection $\Gamma_{\mu}\left(x\right)$, and obtain $\gamma^{\mu}\Gamma_{\mu}\left(x\right)=\gamma^{1}/2\rho$. In this way, the Dirac equation (\ref{2.9}) becomes  
\begin{eqnarray}
i\frac{\partial\psi}{\partial t}=m\hat{\beta}\psi-i\hat{\alpha}^{1}\left[\frac{\partial}{\partial\rho}+\frac{1}{2\rho}+m\omega_{0}\rho\,\hat{\beta}\right]\psi-i\frac{\hat{\alpha}^{2}}{\eta\rho}\frac{\partial\psi}{\partial\varphi}-i\hat{\alpha}^{3}\,\frac{\partial\psi}{\partial z}+i\omega\,\frac{\partial\psi}{\partial\varphi}.
\label{2.11}
\end{eqnarray}

We can see in (\ref{2.11}) that the Dirac Hamiltonian commutes with the $z$-component of the total angular momentum operator $\hat{J}_{z}=-i\partial_{\varphi}$ \cite{schu}, and the $z$-component of the momentum $\hat{p}_{z}=-i\partial_{z}$. Thus, we can take the solutions of Eq. (\ref{2.11}) in the terms of the eigenvalues of the operators $\hat{J}_{z}=-i\partial_{\varphi}$ and $\hat{p}_{z}=-i\partial_{z}$:
\begin{eqnarray}
\psi=\,e^{-i\mathcal{E}\,t}\,e^{ij\varphi}\,e^{ikz}\,\left(
\begin{array}{c}
\phi\left(\rho\right)\\
\chi\left(\rho\right)\\	
\end{array}\right),
\label{2.12}
\end{eqnarray}
where $j=l+\frac{1}{2}$, $l=0,\pm1,\pm2,...$, and $k$ is a constant. We also have in (\ref{2.12}) that $\phi\left(\rho\right)=\left(\phi_{+}\,\,\,\phi_{-}\right)^{T}$ and $\chi\left(\rho\right)=\left(\chi_{+}\,\,\,\chi_{-}\right)^{T}$ are two-spinors, with $\sigma^{3}\,\phi_{+}=\phi_{+}$, $\sigma^{3}\,\phi_{-}=-\phi_{-}$, and the same for $\chi_{\pm}$. Our interest in this work is to study a planar system, therefore we consider $k=0$ from now on. Then, substituting (\ref{2.12}) into (\ref{2.11}), we obtain two coupled equations for $\phi\left(\rho\right)$ and $\chi\left(\rho\right)$, where the first coupled equation is
\begin{eqnarray}
\left[\mathcal{E}-m+\omega\left(l+1/2\right)\right]\phi=-i\sigma^{1}\left[\frac{\partial}{\partial\rho}+\frac{1}{2\rho}-m\omega_{0}\rho\right]\chi+\frac{\sigma^{2}}{\eta\rho}\left(l+1/2\right)\,\chi,
\label{2.13}
\end{eqnarray}
while the second coupled equation is
\begin{eqnarray}
\left[\mathcal{E}+m+\omega\left(l+1/2\right)\right]\chi=-i\sigma^{1}\left[\frac{\partial}{\partial\rho}+\frac{1}{2\rho}+m\omega_{0}\rho\right]\phi+\frac{\sigma^{2}}{\eta\rho}\left(l+1/2\right)\,\phi,
\label{2.14}
\end{eqnarray}

In order to solve the coupled equations (\ref{2.13}) and (\ref{2.14}), we can eliminate $\chi$ from Eq. (\ref{2.14}) and substitute into (\ref{2.13}), then, we obtain two non-coupled second order differential equations for $\phi_{+}$ and $\phi_{-}$. In the following, we write these two non-coupled differential equations in a compact form by labeling the components $\phi_{+}$ and $\phi_{-}$ as $\phi_{s}$, where $s=\pm1$ and $\sigma^{3}\phi_{s}=\pm\phi_{s}=s\phi_{s}$. Thus, the non-coupled differential equation for $\phi_{+}$ and $\phi_{-}$ are
\begin{eqnarray}
\frac{d^{2}\phi_{s}}{d\rho^{2}}+\frac{1}{\rho}\,\frac{d\phi_{s}}{d\rho}-\frac{\zeta^{2}_{s}}{\eta^{2}\rho^{2}}\,\phi_{s}-m^{2}\omega^{2}_{0}\rho^{2}\,\phi_{s}+\tau_{s}\,\phi_{s}=0,
\label{2.15}
\end{eqnarray}
where we have defined in (\ref{2.15}) the effective angular moment $\zeta_{s}=l+\frac{1}{2}\left(1-s\right)+\frac{s}{2}\left(1-\eta\right)$ , and the parameter $\tau_{s}=\left[\mathcal{E}+\omega\left(l+1/2\right)\right]^{2}-m^{2}+2s\,m\omega_{0}\frac{\zeta_{s}}{\eta}+2m\omega_{0}$. 
Next, let us make a coordinate transformation given by $\mu=m\omega_{0}\rho^{2}$. Thus, we obtain the following second order differential equation:
\begin{eqnarray}
\mu\,\frac{d^{2}\phi_{s}}{d\mu^{2}}+\frac{d\phi_{s}}{d\mu}-\frac{\zeta^{2}_{s}}{4\eta^{2}\mu}\,\phi_{s}-\frac{\mu}{4}\,\phi_{s}+\frac{\tau_{s}}{4m\omega_{0}}\,\phi_{s}=0.
\label{2.17}
\end{eqnarray}

In order to have a regular solution at the origin, the solution for the equation (\ref{2.17}) has the form $\phi_{s}\left(\mu\right)=e^{-\frac{\mu}{2}}\,\mu^{\frac{\left|\zeta_{s}\right|}{2\eta}}\,F_{s}\left(\mu\right)$. Thus, substituting this solution into (\ref{2.17}), we obtain
\begin{eqnarray}
\mu\frac{d^{2}F_{s}}{d\mu^{2}}+\left[\frac{\left|\zeta_{s}\right|}{\eta}+1-\mu\right]\frac{dF_{s}}{d\mu}+\left[\frac{\tau_{s}}{4m\omega_{0}}-\frac{\left|\zeta_{s}\right|}{2\eta}-\frac{1}{2}\right]F_{s}=0,
\label{2.18}
\end{eqnarray}
which is the confluent hypergeometric equation or the Kummer equation \cite{abra}. The solution of Eq. (\ref{2.18}) regular at the origin is called the Kummer function of first kind, which is given by $F_{s}\left(\mu\right)=\,_{1}F_{1}\left(\frac{\left|\zeta_{s}\right|}{2\eta}+\frac{1}{2}-\frac{\tau_{s}}{4m\omega_{0}},\frac{\left|\zeta_{s}\right|}{\eta}+1;\mu\right)$. From the solution of Eq. (\ref{2.18}), let us obtain the general form of the Dirac spinors for positive-energy solutions. In order to obtain the appropriate solutions of the Dirac equation (\ref{2.11}), we must solve the system of coupled equations given eqs. (\ref{2.13}) and (\ref{2.14}). By writing $\phi_{s}\left(\rho\right)=e^{-\frac{m\omega_{0}\rho^{2}}{2}}\,\left(m\omega_{0}\rho^{2}\right)^{\frac{\left|\zeta_{s}\right|}{2\eta}}\,_{1}F_{1}\left(\frac{\left|\zeta_{s}\right|}{2\eta}+\frac{1}{2}-\frac{\tau_{s}}{4m\omega_{0}},\frac{\left|\zeta_{s}\right|}{\eta}+1;m\omega_{0}\,\rho^{2}\right)$, and substituting $\phi_{s}$ into (\ref{2.14}), we can obtain the two-spinor $\chi_{s}$. Hence, the positive-energy solutions of the Dirac equation (\ref{2.11}) corresponding to the the parallel components of the Dirac spinor to the $z$-axis of the spacetime becomes
\begin{eqnarray}
\psi_{+}&=&g_{+}\,_{1}F_{1}\left(\frac{\left|\zeta_{+}\right|}{2\eta}+\frac{1}{2}-\frac{\tau_{+}}{4m\omega_{0}},\frac{\left|\zeta_{+}\right|}{\eta}+1;m\omega_{0}\rho^{2}\right)\left(
\begin{array}{c}
1	\\
0 \\
0 \\
-i\frac{\left|\zeta_{+}\right|}{\eta\rho}+i\frac{\zeta_{+}}{\eta\rho}\\
\end{array}\right)\nonumber\\
[-2mm]\label{2.18a}\\[-2mm]
&-&i\,g_{+}\,2m\omega_{0}\rho\,\frac{\left[\frac{\left|\zeta_{+}\right|}{2\eta}+\frac{1}{2}-\frac{\tau_{+}}{4m\omega_{0}}\right]}{\left(\frac{\left|\zeta_{+}\right|}{\eta}+1\right)}\,_{1}F_{1}\left(\frac{\left|\zeta_{+}\right|}{2\eta}+\frac{3}{2}-\frac{\tau_{+}}{4m\omega_{0}},\frac{\left|\zeta_{+}\right|}{\eta}+2;m\omega_{0}\rho^{2}\right)\left(
\begin{array}{c}
0\\
0\\
0\\
1\\	
\end{array}\right),\nonumber
\end{eqnarray}
while the positive-energy solutions of the Dirac equation (\ref{2.11}) corresponding to the the antiparallel components of the Dirac spinor to the $z$-axis of the spacetime becomes
\begin{eqnarray}
\psi_{-}&=&g_{-}\,_{1}F_{1}\left(\frac{\left|\zeta_{-}\right|}{2\eta}+\frac{1}{2}-\frac{\tau_{-}}{4m\omega_{0}},\frac{\left|\zeta_{-}\right|}{\eta}+1;m\omega_{0}\rho^{2}\right)\left(
\begin{array}{c}
0	\\
1 \\
-i\frac{\left|\zeta_{-}\right|}{\eta\rho}-i\frac{\zeta_{-}}{\eta\rho}\\
0 \\
\end{array}\right)\nonumber\\
[-2mm]\label{2.18b}\\[-2mm]
&-&i\,g_{-}\,2m\omega_{0}\rho\,\frac{\left[\frac{\left|\zeta_{-}\right|}{2\eta}+\frac{1}{2}-\frac{\tau_{-}}{4m\omega_{0}}\right]}{\left(\frac{\left|\zeta_{-}\right|}{\eta}+1\right)}\,_{1}F_{1}\left(\frac{\left|\zeta_{-}\right|}{2\eta}+\frac{3}{2}-\frac{\tau_{-}}{4m\omega_{0}},\frac{\left|\zeta_{-}\right|}{\eta}+2;m\omega_{0}\rho^{2}\right)\left(
\begin{array}{c}
0\\
0\\
1\\
0\\	
\end{array}\right),\nonumber
\end{eqnarray}
where we define the parameter $g_{\pm}=g_{s}$ in (\ref{2.18a}) and (\ref{2.18b}) as
\begin{eqnarray}
g_{s}=C\,e^{-i\mathcal{E}t}\,e^{i\left(l+\frac{1}{2}\right)\varphi}\,\frac{\left(m\omega_{0}\rho^{2}\right)^{\frac{\left|\zeta_{s}\right|}{2\eta}}\,e^{-\frac{m\omega_{0}}{2}\rho^{2}}}{\left[\mathcal{E}+m+\omega\left(l+1/2\right)\right]}.
\label{2.18c}
\end{eqnarray}

From now on, let us discuss the influence of noninertial effects on the bound states of the Dirac oscillator in the cosmic string background. It is well known in the literature that the radial part of the wave function becomes finite everywhere when the parameter $A=\frac{\left|\zeta_{s}\right|}{2\eta}+\frac{1}{2}-\frac{\tau_{s}}{4m\omega_{0}}$ of the confluent hypergeometric function is equal to a nonpositive integer number, making the confluent hypergeometric series to be a polynomial of degree $n$ \cite{landau2}. Thus, in order to have a wave function being normalized inside the range $0\,<\,\rho\,<\,1/\omega\eta$, we assume that $\sqrt{m\omega_{0}}\ll\omega\eta$. This assumption makes the the amplitude of probability being very small for values where $\rho\,>\,1/\omega\eta$, because we have that $\mu=m\omega_{0}\rho^{2}\ll1$ when $\rho\rightarrow1/\omega\eta$. Therefore, without loss of generality, we consider the wave function being normalized in the range $0\,<\,\rho\,<\,1/\omega\eta$, since $\phi_{s}\left(\mu\right)\approx0$ when $\rho\rightarrow1/\omega\eta$. In this way, by assuming $\sqrt{m\omega_{0}}\ll\omega\eta$ and imposing $\frac{\left|\zeta_{s}\right|}{2\eta}+\frac{1}{2}-\frac{\tau_{s}}{4m\omega_{0}}=-n$ (where $n=0,1, 2,...$), the relativistic energy levels are
\begin{eqnarray}
\mathcal{E}_{n,\,l}=\sqrt{m^{2}+4m\omega_{0}\left[n+\frac{\left|\zeta_{s}\right|}{2\eta}-s\,\frac{\zeta_{s}}{2\eta}\right]}-\omega\left[l+1/2\right].
\label{2.19}
\end{eqnarray}
with $\zeta_{s}=l+\frac{1}{2}\left(1-s\right)+\frac{s}{2}\left(1-\eta\right)$ being to the effective angular momentum defined previously. The relativistic energy levels (\ref{2.19}) correspond to the relativistic spectrum of energy of the Dirac oscillator under the influence of the noninertial effects of the Fermi-Walker reference frame in the cosmic string background. We have that the bound states of the Dirac oscillator in this noninertial system are obtained by assuming $\sqrt{m\omega_{0}}\ll\omega\eta$. Without assuming $\sqrt{m\omega_{0}}\ll\omega\eta$, we cannot consider the amplitude of probability of finding the Dirac neutral particle in the non-physical region of the spacetime being null. Moreover, we have that curvature effects on the bound states break the degeneracy of the relativistic energy levels of the Dirac oscillator (\ref{2.19}). By taking the limit $\eta\rightarrow1$, we recover the spectrum of energy of the Dirac oscillator in the Minkowski spacetime under the influence of the noninertial effects of the Fermi-Walker reference frame. From the influence of the noninertial effects on the Dirac oscillator, we also have the presence of the coupling between the quantum number $l$ and the angular velocity $\omega$ in the relativistic energy levels given by the last term of (\ref{2.19}).

Next, let us discuss the nonrelativistic limit of the energy levels (\ref{2.19}). The nonrelativistic limit of the energy levels (\ref{2.19}) can be obtained by applying the Taylor expansion up to the first order terms. In this way, the nonrelativistic limit of the energy levels (\ref{2.19}) becomes
\begin{eqnarray}
\mathcal{E}_{n\,l}\approx m+2\omega_{0}\left[n+\frac{\left|\zeta_{s}\right|}{2\eta}-s\,\frac{\zeta_{s}}{2\eta}\right]-\omega\left[l+\frac{1}{2}\right],
\label{2.19a}
\end{eqnarray}
where the first term of the nonrelativistic energy levels (\ref{2.19a}) corresponds to the rest energy of the quantum particle. The remaining terms of (\ref{2.19a}) correspond to the energy levels of a harmonic oscillator under the influence of noninertial effects and the topology of a disclination \cite{kat}. Furthermore, the energy levels (\ref{2.19a}) can be viewed as the spectrum of energy of bound states corresponding to the confinement of a nonrelativistic Dirac neutral particle to a quantum dot described by the Tan-Inkson model \cite{tan} induced by noninertial effects \cite{b2}. The Tan-Inkson model for a quantum dot \cite{tan} is characterized by a confining potential given by $V\left(\rho\right)=a_{2}\,\rho^{2}$, where the spectrum of energy is non-parabolic (proportional to $n$) with a high degeneracy. Note that, in the relativistic radial equation (\ref{2.15}), we have that the Dirac oscillator coupling yields a term proportional to $\rho^{2}$ where the control parameter $a_{2}$ of the Tan-Inkson model is given by $a_{2}=m^{2}\omega_{0}^{2}$, which allows us to make an analogy between the models. However, the non-parabolic spectrum of energy given in (\ref{2.19a}) is given by the noninertial effects when we assume $\sqrt{m\omega_{0}}\ll\omega\eta$. As we have seen above, without the assumption $\sqrt{m\omega_{0}}\ll\omega\eta$, we cannot consider the wave function being normalized inside the physical region of the spacetime. On the other hand, we have that the effects of curvature on the nonrelativistic energy levels (\ref{2.19a}) breaks the degeneracy of the energy levels of bound states as in the Landau quantization for neutral particles \cite{bf11,b}, and in the confinement of a neutral particle to a quantum dot via noninertial effects \cite{b2}. Furthermore, we have that the coupling between the quantum number $l$ and the angular velocity $\omega$ which corresponds to the Page-Werner {\it et al.} term \cite{r1,r2}.

In the following, we wish to make a new discussion without assuming that $\sqrt{m\omega_{0}}\ll\omega\eta$. In this new case, by imposing the condition where the confluent hypergeometric series becomes a polynomial of degree $n$, we obtain a radial wave function finite everywhere (including the non-physical region $\rho\geq1/\omega\eta$). In order that a normalized wave function can be obtained in this new case, {\it we first consider $m\omega_{0}$ is quite small}. The assumption $m\omega_{0}$ is small allows us to consider a fixed radius $\rho_{0}=1/\omega\eta$ in such a way, by taking a fixed value for the parameter $B=\frac{\left|\zeta_{s}\right|}{\eta}+1$ of the confluent hypergeometric function, that we can consider the parameter $A=\frac{\left|\zeta_{s}\right|}{2\eta}+\frac{1}{2}-\frac{\tau_{s}}{4m\omega_{0}}$ of the confluent hypergeometric function being large. In this way, we can write the confluent hypergeometric function in terms of the Bessel function of first kind in the form \cite{abra}:
\begin{eqnarray}
_{1}F_{1}\left(A,B,\mu_{0}=m\omega_{0}\,\rho^{2}_{0}\right)&\approx&\frac{\Gamma\left(B\right)}{\sqrt{\pi}}\,e^{\frac{\mu_{0}}{2}}\left(\frac{B\mu_{0}}{2}-A\mu_{0}\right)^{\frac{1-B}{2}}\nonumber\\
&\times&\cos\left(\sqrt{2B\mu_{0}-4A\mu_{0}}-\frac{B\pi}{2}+\frac{\pi}{4}\right),
\label{2.20}
\end{eqnarray}
where $\Gamma\left(B\right)$ is the gamma function. Hence, our last step in order to obtain a normalized wave function in the range $0\,<\,\rho\,<\,1/\omega\eta$ is to impose that the radial wave function vanishes at $\rho\rightarrow1/\omega\eta$, that is,  
\begin{eqnarray}
\phi_{s}\left(\mu_{0}\right)=\phi_{s}\left(m\omega_{0}\rho_{0}^{2}\right)=0,
\label{2.22}
\end{eqnarray}
where $\rho_{0}=1/\omega\eta$. In this way, by writing the radial wave function $\phi_{s}\left(\mu\right)=e^{-\frac{\mu}{2}}\,\mu^{\frac{\left|\zeta_{s}\right|}{2\eta}}\,F_{s}\left(\mu\right)$ in terms of (\ref{2.20}) and by applying the boundary condition (\ref{2.22}), we have that the relativistic energy levels of the Dirac oscillator in the Fermi-Walker reference frame become
\begin{eqnarray}
\mathcal{E}_{n,\,l}\approx\sqrt{m^{2}+\frac{1}{\rho_{0}^{2}}\left[n\pi+\frac{\zeta_{s}\pi}{2\eta}+\frac{3\pi}{4}\right]^{2}-2s\,m\omega_{0}\frac{\zeta_{s}}{\eta}-2m\omega_{0}}-\omega\left[l+\frac{1}{2}\right].
\label{2.23}
\end{eqnarray} 

Hence, we have that the conditions of vanishing the radial wave function at $\rho\rightarrow1/\omega\eta$ and $m\omega_{0}\ll1$ yield both a normalized radial wave function inside the physical region of the spacetime, and the relativistic spectrum of energy of the Dirac oscillator under the influence of noninertial effects (\ref{2.23}). But we can see that the relativistic energy levels (\ref{2.23}) differ from the energy levels obtained in (\ref{2.19}) even though both cases result from noninertial effects. This difference arises from the conditions imposed on $\omega_{0}$ ($m\omega_{0}$ is quite small), and from the restriction of the physical region of the spacetime imposed by noninertial effects. We also have that the effects of curvature on the relativistic energy levels (\ref{2.23}) change the degeneracy of the spectrum of energy. Again, by taking the limit $\eta\rightarrow1$, the curvature effects vanish and we recover the spectrum of energy of the Dirac oscillator under the influence of noninertial effects in the Minkowski spacetime.

Now, let us take the nonrelativistic limit of the energy levels (\ref{2.23}). The nonrelativistic limit of the energy levels (\ref{2.23}) can also be obtained by applying the Taylor expansion up to the first order terms. Thus, the nonrelativistic limit of the energy levels (\ref{2.23}) are
\begin{eqnarray}
\mathcal{E}_{n\,l}\approx m+\frac{1}{2m\rho_{0}}\left[n\pi+\frac{\zeta_{s}\pi}{2\eta}+\frac{3\pi}{4}\right]^{2}-\omega_{0}\left[1+s\frac{\zeta_{s}}{\eta}\right]-\omega\left[l+\frac{1}{2}\right],
\label{2.24}
\end{eqnarray}
where the first term of the nonrelativistic energy levels (\ref{2.24}) also corresponds to the rest energy of the quantum particle. Note that the energy levels (\ref{2.24}) corresponds to the bound states of a nonrelativistic Dirac particle confined to to the region of the spacetime $0\,<\,\rho\,<\,1/\omega\eta$. This confinement to the physical region of the spacetime is analogous to having a neutral particle confined to a quantum dot described by a hard-wall confining potential \cite{dot,dot2,dot3}. Recently, a hard-wall confining potential has been used in studies of the confinement of quantum particles to a magnetic quantum dot \cite{dot2}, quantum antidots for Landau states \cite{dot}, and in the confinement of a neutral particle with a permanent magnetic dipole moment interacting with a radial electric field to a quantum dot \cite{bf20}. Comparing the result (\ref{2.24}) with previous studies \cite{b2} where the analogous confinement of a neutral particle to a quantum dot is given by imposing a condition on the induced fields $\left(\mu\lambda\ll\omega\right)$, we have in the present case that the geometry of the spacetime plays the role of a hard-wall confining potential due to the presence of noninertial effects that restricts the physical region of the spacetime. Moreover, we have that the energy levels (\ref{2.24}) are proportional to $n^{2}$ in contrast to the previous result (\ref{2.19a}), where the energy levels are proportional to $n$ in analogous way to the Tan-Inkson model for a quantum dot \cite{tan,b2}. Finally, note that the last term of (\ref{2.23}) corresponds to the Page-Werner {\it et al.} term \cite{r1,r2}, and the breaking of the degeneracy of the energy levels given by the topology of the defect.

\section{conclusions}

In this work, we have studied the influence of curvature and noninertial effects on the Dirac oscillator by using the spinor theory in curved spacetime \cite{weinberg}. We have shown that the influence of the noninertial effects of the Fermi-Walker reference frame allows us to obtain two distinct radial solutions of the Dirac equation and, consequently, yields distinct relativistic spectra of energy. 

We have seen, by assuming $\sqrt{m\omega_{0}}\ll\omega\eta$, that the amplitude of the wave function of the Dirac neutral particle becomes very small for values where $\rho\,>\,1/\omega\eta$, thus, we can consider the wave function being normalized inside the physical region of the spacetime defined by the range $0\,<\,\rho\,<\,1/\omega\eta$. As a consequence of this assumption, we have obtained that the relativistic spectrum of energy of the Dirac oscillator under the influence of the noninertial effects of the Fermi-Walker reference frame is analogous to the spectrum of the energy of the confinement of a relativistic Dirac neutral particle to a quantum dot described by a parabolic potential \cite{b4,bf30}. Moreover, we have seen that the curvature effects on the bound states change the degeneracy of the relativistic energy levels of the Dirac oscillator. On the other hand, by taking the nonrelativistic limit of the relativistic energy levels of the Dirac oscillator (with $\sqrt{m\omega_{0}}\ll\omega\eta$), we have seen that the nonrelativistic energy levels of the bound states are nonparabolic, and corresponding to the analogous case of the confinement of a spin-half neutral particle to a quantum dot described by the Tan-Inkson model \cite{tan} under the influence of noninertial effects \cite{b2}. Furthermore, the effects of curvature on the nonrelativistic energy levels also change the degeneracy of the energy levels of bound states.   

We have also discussed the case where the assumption $\sqrt{m\omega_{0}}\ll\omega\eta$ is not valid anymore. In this case, we have shown that we cannot normalize the radial wave function by imposing that the confluent hypergeometric series becomes a polynomial of degree $n$, because the radial wave function becomes defined both in the physical region of the spacetime and in the non-physical region of the spacetime. In this way, in order that a normalized radial wave function can be obtained inside the physical region the spacetime defined by the range $0\,<\,\rho\,<\,1/\omega\eta$, we have imposed that the radial wave function vanishes at $\rho\rightarrow1/\omega\eta$, and $m\omega_{0}\ll1$. Hence, the conditions given by vanishing the radial wave function at $\rho\rightarrow1/\omega\eta$ and $m\omega_{0}$ being quite small have yielded a normalized radial wave function, and a relativistic spectrum of energy of the Dirac oscillator under the influence of noninertial effects which differs from the energy levels obtained previously by considering the condition $\sqrt{m\omega_{0}}\ll\omega\eta$. Moreover, we have shown that the nonrelativistic limit of the energy levels of the Dirac oscillator under the influence of the noninertial effects of the Fermi-Walker reference frame obtained in this second case also correspond to a analogous confinement of a Dirac neutral particle to a quantum dot described by a hard-wall confining potential as discussed in Refs. \cite{dot2,dot,dot3,bf20}. 

\acknowledgments{We would like to thank CNPq (Conselho Nacional de Desenvolvimento Cient\'ifico e Tecnol\'ogico - Brazil) for financial support.}

\end{document}